\documentclass[prl,aps,twocolumn,superscriptaddress,%showpacs
]{revtex4}
\usepackage{amsmath,amssymb,graphicx}
\usepackage{pxfonts}
\usepackage{graphicx}
\usepackage{color}
\usepackage{float}
\begin{document}

%\date{\today}

\title{Efficient transport kinetics of indirect excitons in van der Waals heterostructure}

\author{Zhiwen~Zhou}
\affiliation{Department of Physics, University of California San Diego, La Jolla, CA 92093, USA}
\author{W.~J.~Brunner}
\affiliation{Department of Physics, University of California San Diego, La Jolla, CA 92093, USA}
\author{E.~A.~Szwed}
\affiliation{Department of Physics, University of California San Diego, La Jolla, CA 92093, USA}
\author{H.~Henstridge}
\affiliation{Department of Physics, University of California San Diego, La Jolla, CA 92093, USA}
\author{L.~H.~Fowler-Gerace}
\affiliation{Department of Physics, University of California San Diego, La Jolla, CA 92093, USA}
\author{L.~V.~Butov} 
\affiliation{Department of Physics, University of California San Diego, La Jolla, CA 92093, USA}

\begin{abstract}
\noindent Exciton transport is fundamental for understanding transport phenomena in bosonic systems and for exploring excitation energy transfer in materials. Spatially indirect excitons (IXs) have long lifetimes allowing them to form quantum bosonic states and travel long distances. Van der Waals heterostructures form a new materials platform for exploring IX transport. Disordered in-plane potentials suppress IX transport due to IX localization and scattering. In this work, we found the efficient IX transport kinetics 
characterized by anomalously high IX mobility. The efficient IX transport kinetics is observed in the presence of in-plane disorder and is consistent with predicted IX superfluidity.
\end{abstract}
\maketitle

Excitons, the bound pairs of electrons and holes, are composite bosons~\cite{Keldysh1968}. Spatially indirect excitons (IXs), also known as interlayer excitons, are formed by electrons and holes in separated layers in a heterostructure (HS)~\cite{Lozovik1976}. Due to the layer separation, the IX lifetimes are orders of magnitude longer than lifetimes of spatially direct excitons (DXs)~\cite{Zrenner1992}. The long lifetimes allow IXs to cool below the temperature of quantum degeneracy and form quantum bosonic states~\cite{High2012}. Furthermore, the long IX lifetimes give an opportunity to achieve long-range exciton transport~\cite{Hagn1995}. Due to these properties, IXs can be used for studying transport phenomena in quantum bosonic states and for achieving efficient excitonic energy transfer in materials.

IX transport is studied in various materials platforms, including GaAs HS~\cite{Hagn1995, Larionov2000, Butov2002, Voros2005, Ivanov2006, Gartner2006, High2008, Remeika2009, Vogele2009, Lazic2010, Alloing2012, Remeika2012, High2013, Lazic2014, Alloing2014, Gorbunov2016, Leonard2018, Leonard2021}, GaN HS~\cite{Chiaruttini2019}, and ZnO HS~\cite{Kuznetsova2015}. Recently, IX transport studies were started in van der Waals (vdW) HS composed of atomically thin layers of transition metal dichalcogenides (TMD). VdW TMD HS offer a unique platform for exploring exciton transport. First, excitons in vdW TMD HS, both DXs~\cite{Ye2014, Chernikov2014, Goryca2019} and IXs~\cite{Fogler2014, Deilmann2018}, have high binding energies reaching hundreds of meV. In comparison, IX binding energies in HS formed from III-V or II-VI semiconductors are significantly lower: e.g. $3-4$~meV in GaAs/AlGaAs HS~\cite{Sivalertporn2012, Szwed2024} and 30~meV in ZnO HS~\cite{Morhain2005}. The high binding energies make IXs in vdW TMD HS stable at room temperature. Furthermore, the predicted superfluidity temperature that can be achieved with excitons is proportional to the exciton binding energy and the high IX binding energies in vdW TMD HS give an opportunity to realize high-temperature superfluidity~\cite{Fogler2014}. Second, IXs in periodic potentials of moir{\'e} superlattices in vdW TMD HS give an opportunity to explore the Bose-Hubbard (BH) physics~\cite{Wu2018, Yu2018, Wu2017, Yu2017, Zhang2017a}. 

Exciton transport is intensively studied both with DXs in TMD monolayers~\cite{Kumar2014, Kulig2018, Cadiz2018, Leon2018, Leon2019, Hao2020, Datta2022} and with IXs in vdW TMD HS~\cite{Rivera2016, Jauregui2019, Unuchek2019a, Unuchek2019, Liu2019, Choi2020, Huang2020, Yuan2020, Li2021, Wang2021, Shanks2022, Sun2022, Tagarelli2023, Rossi2023, Gao2024, Zhang2024, Fowler-Gerace2021, Peng2022, Wietek2024, Troue2023, Cutshall2025}. These studies showed that disordered in-plane potentials, including disordered moir{\'e} potentials, suppress diffusive IX transport due to IX localization and scattering. In particular, even in the case of long IX lifetimes, diffusive IX transport in vdW TMD HS is characterized by low $1/e$ decay distances $d_{1/e}$, up to a few micrometers~\cite{Rivera2016, Jauregui2019, Unuchek2019a, Unuchek2019, Liu2019, Choi2020, Huang2020, Yuan2020, Li2021, Wang2021, Shanks2022, Sun2022, Tagarelli2023, Rossi2023, Gao2024, Zhang2024, Wietek2024}. Recent cw studies showed that $d_{1/e}$ can be significantly longer~\cite{Fowler-Gerace2024, Zhou2024}. However, 
cw studies do not measure the exciton transport kinetics and, therefore, do not distinguish a regular diffusion from a non-diffusive transport and do not determine the transport characteristics such as the IX mobility. In particular, cw studies do not distinguish a linear growth of $d_{1/e}^2$ with time, characteristic of diffusion, from a quadratic growth of $d_{1/e}^2$ with time, characteristic of ballistic transport or drift with a nearly constant velocity. Measurements of transport kinetics $d_{1/e}(t)$ are needed to identify the transport regime and determine the transport characteristics.

In this work, we studied IX transport in a MoSe$_2$/WSe$_2$ vdW HS by time-resolved imaging that directly measures the IX photoluminescence (PL) cloud expansion and reveals the IX transport kinetics. We found the efficient IX transport kinetics 
with a fast and quadratic growth of $d_{1/e}^2$ with time. The efficient IX transport kinetics is characterized by anomalously high IX mobility and is consistent with predicted IX superfluidity.

\vskip 3mm
{\bf Experimental results}

We study MoSe$_2$/WSe$_2$ HS where adjacent MoSe$_2$ monolayer and WSe$_2$ monolayer form the separated electron and hole layers for IXs~\cite{Rivera2015}. The HS details are presented in Supplementary Information (SI).

IXs are optically generated by focused laser excitation with the laser excitation energy resonant to DX in WSe$_2$ HS layer. The laser pulses have step-like rectangular shape with 84~ns pulse width, 0.3~ns edge sharpness, and 300~ns pulse period. The IX transport is measured by time- and spatially-resolved optical imaging. The measurement details are presented in SI.

\begin{figure}
\begin{center}
\includegraphics[width=8.5cm]{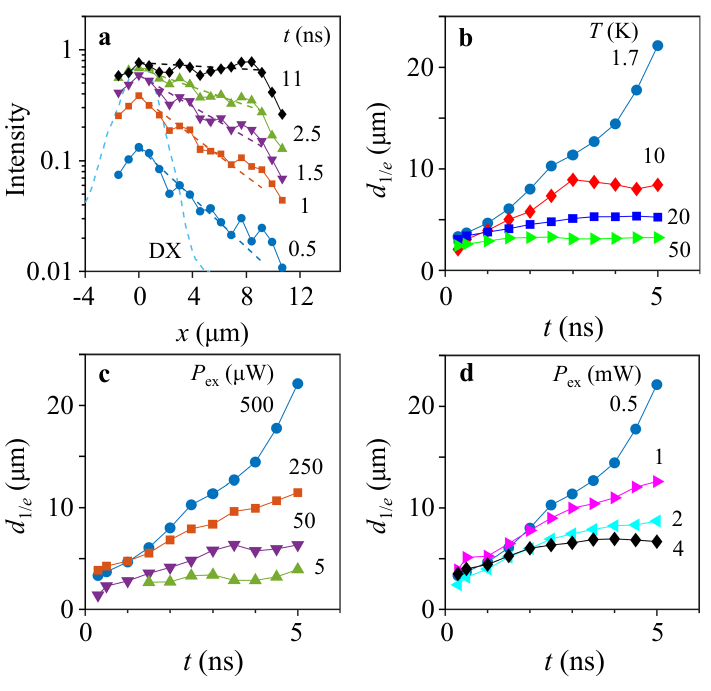}
\caption{Kinetics of IX cloud expansion.
(a) IX PL spatial profiles $I(x)$ vs. time $t$ after the start of laser excitation pulse. IXs are created by the laser excitation. The excitation spot is centered at $x = 0$ and is shown by cyan dotted line that presents the DX PL profile in MoSe$_2$ monolayer and is close to the laser profile for short-range DX transport. The dashed lines show least-squares fits of the IX cloud profiles $I(x, t)$ to exponential decays in the region from the excitation spot to the HS edge, $x=0 - 9$~$\mu$m. 
(b-d) The $1/e$ decay distance of the IX cloud profiles $d_{1/e}$ vs. $t$ for different temperatures $T$ (b) and laser excitation powers $P_{\rm ex}$ (c,d). $P_{\rm ex} = 0.5$~mW in (a,b), $T = 1.7$~K in (a,c,d). 
The IX cloud expands and $d_{1/e}$ increases with time due to IX transport. The efficient IX transport kinetics is observed in a certain range of temperatures and densities, and vanishes at high temperatures (b) and at lower (c) or higher (d) excitation densities.
}
\end{center}
\label{fig:spectra}
\end{figure}

After the onset of laser pulse, the IX cloud expands due to IX transport from the excitation spot. Figure~1a shows the kinetics of the IX signal profiles $I(x, t)$. The spatial extension of IX cloud is characterized by its $1/e$ decay distance $d_{1/e}(t)$ that is determined by least-squares exponential fit to $I(x, t)$ in the HS (Fig.~1a). 

Figures~1b-d show the kinetics of $d_{1/e}(t)$ for different temperatures $T$ and laser excitation powers $P_{\rm ex}$. The enhancement of $d_{1/e}(t)$ with time quantifies the kinetics of the IX cloud expansion due to IX transport. The efficient IX transport kinetics with rapid $d_{1/e}(t)$ enhancement is observed in a certain range of temperatures and densities and vanishes at high temperatures (Fig.~1b) and at lower (Fig.~1c) or higher (Fig.~1d) excitation densities.

\begin{figure*}
\begin{center}
\includegraphics[width=12.5cm]{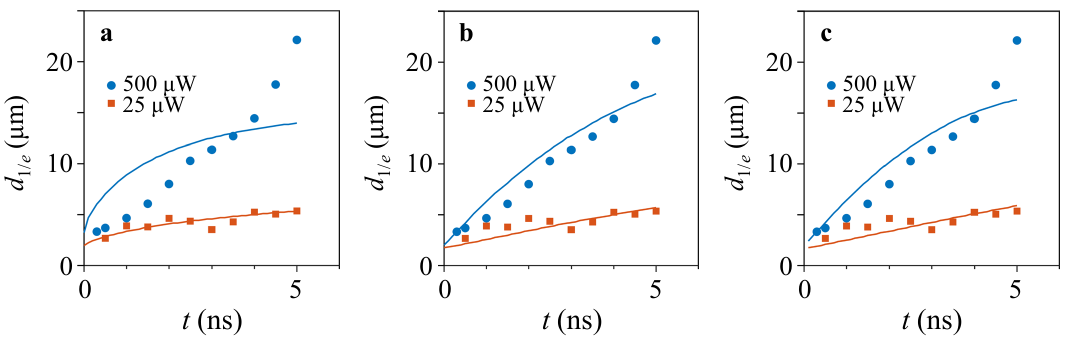}
\caption{Simulations of IX cloud expansion kinetics. 
The measured IX cloud expansion $d_{1/e}(t)$ is shown by blue points for excitation power $P_{\rm ex} = 500$~$\mu$W and by red squares for $P_{\rm ex} = 25$~$\mu$W. The simulated $d_{1/e}(t)$ for diffusion (a), ballistic (b), 
and drift-diffusion (c)
IX transport are shown by lines. In the simulations, the IX diffusivity $D$ for diffusion (a), IX velocity $v$ for ballistic (b), 
and IX mobility $\mu$ for drift-diffusion (c) 
transport are estimated by least-squares fit of the kinetics of exciton cloud expansion $d_{1/e}(t)$ simulated with Eq.~1, Eq.~2, 
and Eq.~3, 
respectively, to the measured $d_{1/e}(t)$. The obtained $D$, $v$, 
and $\mu$ 
are presented in Fig.~3. The laser excitation pulse starts at $t = 0$. $T = 1.7$~K.
}
\end{center}
\label{fig:spectra}
\end{figure*}

\vskip 3mm
{\bf Analysis of experimental data}

For characterization of IX transport kinetics, we use two elemental approaches -- in terms of diffusion (Eq.~1) and ballistic transport (Eq.~2), and an approach including drift (Eq.~3). These approaches attribute quantitative characteristics to IX transport such as diffusivity, velocity, and mobility that enables transport comparison at different parameters. 

The advantage of the first two approaches is their simplicity. Each of them has only one fit parameter -- IX diffusivity $D$ or velocity $v$. However, these two approaches neglect IX drift. The third approach is more complex, however, it includes both the drift due to IX interaction and the drift due to the HS long-scale in-plane potential. Adopting the mean-field model for IX interaction~\cite{Yoshioka1990} allows describing IX transport including drift with one fit parameter -- IX mobility $\mu$. These approaches are outlined below.

Within the first approach, we characterize the measured IX transport by comparing it with diffusion 
described by the diffusion equation for IX density $n$
\begin{eqnarray}
\frac{\partial n}{\partial t} = D \Delta n + \Lambda - \frac{n}{\tau}\,
\end{eqnarray}
where the first term describes IX diffusion with diffusivity $D$, the second IX generation by laser excitation, and the third IX recombination. The IX generation profile $\Lambda(x)$ is given by the DX PL profile in MoSe$_2$ monolayer that is close to the laser profile for the short-range DX transport (Fig.~1a). The IX generation by laser excitation pulse with the rectangular step-like shape starts at $t = 0$. The IX recombination lifetime $\tau$ is obtained from the measurements of the decay kinetics of the spatially integrated IX signal after the excitation pulse end. The measurements of $\tau$ are outlined in SI. The IX diffusivity $D$ is estimated by least-squares fit of the kinetics of exciton cloud expansion $d_{1/e}(t)$ simulated with Eq.~1 to the measured $d_{1/e}(t)$ (Fig.~2a). 

Within the second approach, we characterize the measured IX transport by comparing it 
with ballistic transport described by the equation
\begin{eqnarray}
\frac{\partial n}{\partial t} = - v \nabla n + \Lambda - \frac{n}{\tau}\,
\end{eqnarray}
where the first term describes IX ballistic transport with velocity $v$. The other terms are similar to the corresponding terms in Eq.~1. The IX velocity $v$ is estimated by least-squares fit of the kinetics of exciton cloud expansion $d_{1/e}(t)$ simulated with Eq.~2 to the measured $d_{1/e}(t)$ (Fig.~2b). 

Next, we consider an approach including drift. We start from the effect of IX interaction. The IX out-of-plane dipoles cause repulsive interaction between IXs as shown by the IX studies both in GaAs HS~\cite{Yoshioka1990, Butov1994, Zhu1995, Lozovik1997, Butov1999, Palo2002, Ivanov2002, Leon2003, Schindler2008, Remeika2009, Laikhtman2009, Ivanov2010, Maezono2013, Remeika2015, Dorow2017} and vdW HS~\cite{Rivera2015, Nagler2017, Kremser2020, Sun2022, Fowler-Gerace2024}. As shown in Ref.~\cite{Yoshioka1990}, the repulsive interaction between IXs can be approximated by the mean-field formula for the IX energy shift with density $\delta E = n u_0$, $u_0 = 4 \pi e^2 d_z/\varepsilon$, where $d_z$ is the separation between the electron and hole layers and $\varepsilon$ is the dielectric constant ($d_z \sim 0.6$~nm and $\varepsilon \sim 7.4$ for the HS~\cite{Laturia2018}). This mean-field model~\cite{Yoshioka1990} is commonly used for estimating IX densities both in GaAs HS and vdW HS, e.g. in Refs.~\cite{Larionov2000, Butov2002, High2008, Vogele2009, Alloing2012, Szwed2024, Jauregui2019, Unuchek2019, Sun2022, Zhang2024, Fowler-Gerace2021, Fowler-Gerace2024, Yoshioka1990, Dorow2017, Ivanov2002}. Including correlations lowers $u_0$ in comparison to the mean-field value, leading to a higher estimated $n$ for a given $\delta E$~\cite{Schindler2008, Remeika2009, Laikhtman2009, Ivanov2010, Remeika2015, Steinhoff2025}. The IX repulsive interaction introduces interaction-induced drift from the origin, adding the term $\nabla \left[ \mu n \nabla (n u_0) \right]$ to 
Eq.~1, where $\mu$ is the IX mobility~\cite{Ivanov2002}. 

We also include IX drift due to the in-plane potential energy landscape in the HS, $U$. The landscape originates from the variation of the local IX environment in the HS and is outlined in SI. This drift adds  
$\nabla \left[ \mu n \nabla U \right]$ to Eq.~1~\cite{Ivanov2002}. 

The third approach to characterize the IX transport kinetics includes both the interaction-induced drift and the drift due to the potential energy landscape in the HS and describes IX transport by the equation
\begin{eqnarray}
\frac{\partial n}{\partial t} = \nabla \left[ D^{\prime} \nabla n + \mu n \nabla (n u_0 + U) \right] + \Lambda - \frac{n}{\tau}\,
\end{eqnarray}
The second term in the square brackets describes the drift due to interaction and in-plane potential energy landscape outlined above. The other terms are similar to the corresponding terms in Eq.~1 ($D$ and $D^{\prime}$ are used in Eq.~1 and 3 to avoid confusion). We use the Einstein relationship $D^{\prime} = \mu k_B T$ ($k_{\rm B}$ is the Boltzmann constant) and estimate IX mobility $\mu$ by least-squares fit of the kinetics of exciton cloud expansion $d_{1/e}(t)$ simulated with Eq.~3 to the measured $d_{1/e}(t)$ (Fig.~2c). 

For the HS geometry elongated along the $x$ direction, as outlined in SI, the IX transport is measured along $x$ and is approximated in the simulations as one dimensional. 
$U(x)$ drops $\sim 10$~meV on $\sim 10$~$\mu$m, as outlined in SI, and is approximated by the energy gradient 1~meV/$\mu$m. The obtained estimates for the IX diffusivity $D$ within the diffusion approach, for the IX velocity $v$ within the ballistic approach, and for the IX mobility $\mu$ within the drift-diffusion approach are presented in Fig.~3.

The simulations for diffusion transport (Fig.~2a) show nearly linear growth of $d_{1/e}^2$ with time (nearly square-root-power growth of $d_{1/e}$ with time). The simulations for ballistic transport (Fig.~2b) and the simulations for drift-diffusion transport (Fig.~2c) show nearly quadratic growth of $d_{1/e}^2$ with time (nearly linear growth of $d_{1/e}$ with time). There are deviations from linear growth of $d_{1/e}^2(t)$ for diffusion and from quadratic growth of $d_{1/e}^2(t)$ for both ballistic and drift-diffusion transport. In particular, exciton recombination makes the growth slower. 

All approaches show similar dependence of IX transport kinetics on parameters: The efficient IX transport kinetics characterized by high $D$, $v$, and $\mu$ (Fig.~3) is observed in a certain range of temperatures and excitation densities and vanishes at high temperatures and at lower or higher excitation densities. The considered approaches quantify the temperature and excitation density dependence of the rapid $d_{1/e}(t)$ enhancement observed in the experiment (Figs.~1b-d).

\begin{figure*}
\begin{center}
\includegraphics[width=17.9cm]{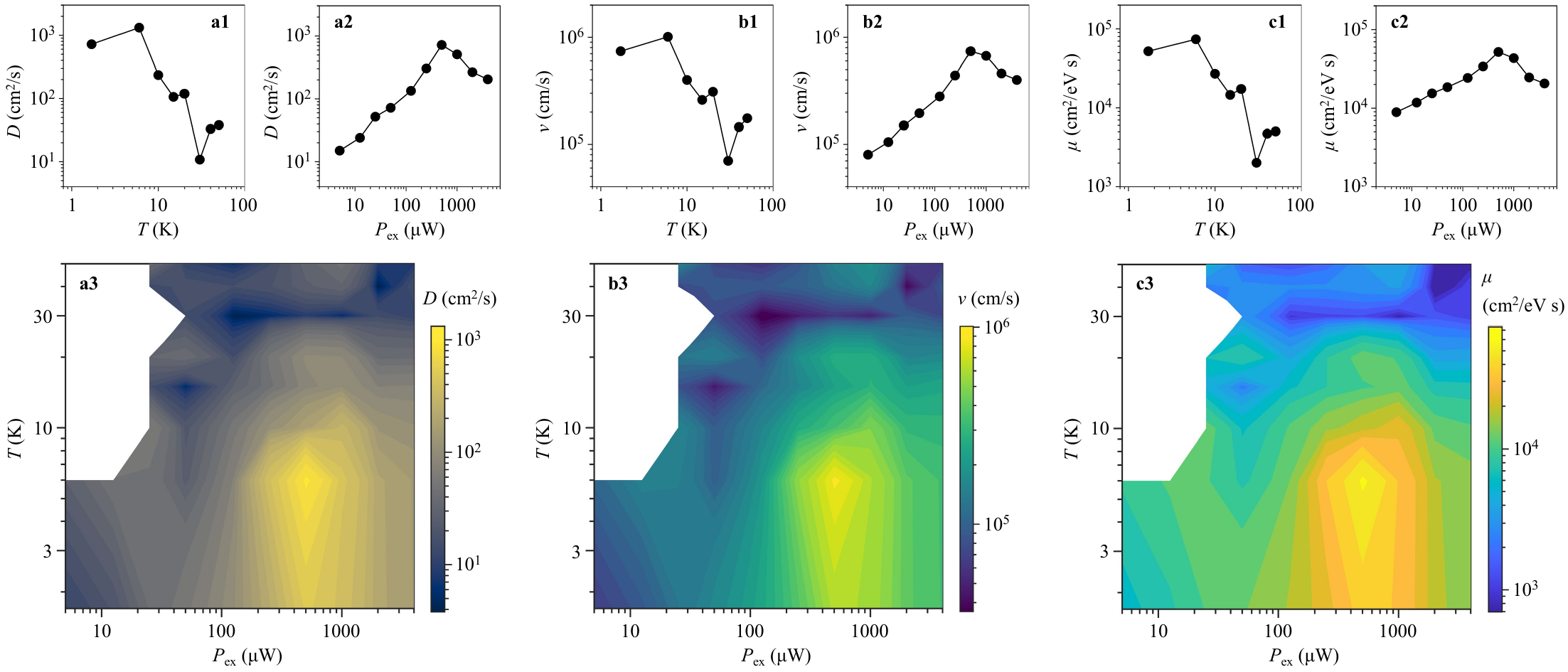}
\caption{Characterization of 
temperature and excitation power dependence of IX transport kinetics: (a) Diffusion approach, (b) ballistic approach, 
and (c) drift-diffusion approach. 
The IX diffusivity $D$ in (a), the IX velocity $v$ in (b), 
and the IX mobility $\mu$ in (c) 
is estimated by least-squares fit of the kinetics of exciton cloud expansion $d_{1/e}(t)$ simulated with Eq.~1 in (a), Eq.~2 in (b), 
and Eq.~3 in (c), 
respectively, to the measured $d_{1/e}(t)$. 
(a): The estimated $D$ vs. temperature $T$ (a1), excitation power $P_{\rm ex}$ (a2), and $T$ and $P_{\rm ex}$ (a3). 
(b): The estimated $v$ vs. $T$ (b1), $P_{\rm ex}$ (b2), and $T$ and $P_{\rm ex}$ (b3).
(c): The estimated $\mu$ vs. $T$ (c1), $P_{\rm ex}$ (c2), and $T$ and $P_{\rm ex}$ (c3). 
$P_{\rm ex} = 0.5$~mW (a1, b1, c1), $T = 1.7$~K (a2, b2, c2).
}
\end{center}
\label{fig:spectra}
\end{figure*}

\vskip 3mm
{\bf Discussion}

{\it Anomalously fast $d_{1/e}(t)$ growth}.
IX transport kinetics is intensively studied in vdW heterostructures~\cite{Yuan2020, Li2021, Wang2021, Sun2022, Tagarelli2023, Wietek2024}. The comparison of the data $d_{1/e}(t)$ in different studies of IX transport kinetics in vdW HS is shown in Fig.~S12 in SI. Figure~S12 shows that $d_{1/e}(t)$ grows significantly faster for the efficient IX transport in comparison to IX transport in earlier studies of vdW HS.

{\it $d_{1/e}(t)$ growth kinetics.}
In earlier studies, IX diffusion or subdiffusion with $d_{1/e}^2$ growing linearly or sublinearly with time was measured~\cite{Yuan2020, Li2021, Wang2021, Sun2022, Tagarelli2023, Wietek2024} 
(Fig.~S12b). In contrast, for the efficient IX transport, $d_{1/e}^2$ grows nearly quadratically with time (Fig.~S12a) or, equivalently, $d_{1/e}$ grows nearly linearly with time (Fig.~2). The nearly linear growth of transport distance $d_{1/e}$ with time corresponds to transport with a nearly constant velocity that is characteristic of ballistic transport or drift (Figs.~2b,c).

The anomalously fast growth of $d_{1/e}$ (Fig.~S12) gives the anomalously high IX mobility $\mu$ (Fig.~3c) and, in turn, the anomalously high IX mean free time $\tau_{\rm m} \sim \mu m$ in the efficient IX transport regime ($m$ is the IX mass in the HS, close to the free electron mass). The estimated IX mean free time for the efficient IX transport kinetics is significantly higher than in earlier studies (see SI). 

The high $\tau_{\rm m}$ corresponds to the high mean free path $l_{\rm m} \propto \tau_{\rm m}$.
The mean free path gives the range of ballistic transport and the high $l_{\rm m}$ for the efficient IX transport kinetics indicates long-range ballistic transport.

The anomalously high IX mobility indicates suppressed IX scattering. IXs move in the HS disorder potential with the amplitude presented by the IX PL linewidth ca. 10~meV (see SI). The realization of the anomalously high IX mobility in the presence of the substantial scattering disorder is qualitatively consistent with the suppression of scattering due to superfluidity predicted for IXs in vdW HS (ref.~\cite{Fogler2014} and refs. therein).

{\it Comparison of IX transport with diffusive transport.}
For 
diffusive transport comprising both diffusion and drift-diffusion 
in the HS in-plane disorder potential, IX transport increases both with temperature, due to IX delocalization from local minima of the disorder potential, and with density, due to IX screening of the disorder potential 
and enhanced interaction-induced drift from the origin~\cite{Ivanov2002, Ivanov2006, Remeika2009}. However, for the efficient IX transport kinetics, a qualitatively different behavior is observed: the efficient IX transport vanishes with temperature and vanishes with density at high densities (Fig.~3), indicating that the efficient IX transport kinetics is inconsistent with the diffusive transport. 

Furthermore, both the estimated amplitude of the moir{\'e} potential~\cite{Wu2018, Yu2018, Wu2017, Yu2017, Zhang2017a} and the disorder amplitude 
outlined above 
are high. They are significantly higher than the thermal energy at the temperatures where the efficient IX transport kinetics is observed ($k_{\rm B}T \lesssim 1$~meV, Fig.~3) so the IX thermal energy is insufficient for effective IX delocalization from the local minima of the in-plane potential for the diffusive IX transport~\cite{Ivanov2002}. They are also significantly higher than the IX interaction energy at the densities where the efficient IX transport kinetics is observed (the IX energy shift due to the interaction $\sim 3$~meV for the excitation densities corresponding to the efficient IX transport kinetics, see SI) so the IX interaction energy is insufficient for effective screening of the in-plane potential for the diffusive IX transport~\cite{Remeika2009, Remeika2012}. Therefore, the efficient IX transport kinetics 
characterized by the anomalously high IX mobility, despite the in-plane disorder, 
is inconsistent with the diffusive transport.

{\it Comparison of IX transport with BH theory.}
The efficient IX transport kinetics characterized by the anomalously high IX mobility, despite the in-plane disorder, is qualitatively consistent with the theoretically predicted superfluidity. 
The BH model predicts the superfluid phase for the number of bosons per lattice site $N \sim 1/2$ and the insulating phase, such as the Mott insulator and the Bose glass, for $N \sim 0$ and $N \sim 1$~\cite{Fisher1989}. The IX transport enhancement followed by the suppression with density (Fig.~3) is qualitatively consistent with this prediction. The efficient IX transport kinetics vanishes above $T \sim 10$~K (Fig.~3). This is also qualitatively consistent with the BH theory prediction for superfluidity. A comparison with the BH model is outlined in SI.

In summary, we found in vdW TMD HS the efficient IX transport kinetics 
characterized by anomalously high IX mobility and
consistent with predicted IX superfluidity.

\vskip 3mm
{\bf Acknowledgements}
We thank M.M.~Fogler, B.~Vermilyea, and A.H.~MacDonald for discussions and A.K.~Geim for teaching us manufacturing TMD HS. The studies were supported by the Department of Energy, Office of Basic Energy Sciences, under award DE-FG02-07ER46449. The HS manufacturing and data analysis were supported by NSF Grants 1905478 and 2516006.

\end{document}

%\date{\today}

\title{Supporting Information for

Efficient transport kinetics of indirect excitons in van der Waals heterostructure}

\author{Zhiwen~Zhou}
\affiliation{Department of Physics, University of California San Diego, La Jolla, CA 92093, USA}
\author{W.~J.~Brunner}
\affiliation{Department of Physics, University of California San Diego, La Jolla, CA 92093, USA}
\author{E.~A.~Szwed}
\affiliation{Department of Physics, University of California San Diego, La Jolla, CA 92093, USA}
\author{H.~Henstridge}
\affiliation{Department of Physics, University of California San Diego, La Jolla, CA 92093, USA}
\author{L.~H.~Fowler-Gerace}
\affiliation{Department of Physics, University of California San Diego, La Jolla, CA 92093, USA}
\author{L.~V.~Butov} 
\affiliation{Department of Physics, University of California San Diego, La Jolla, CA 92093, USA}

\begin{abstract}
\end{abstract}
\maketitle

\renewcommand*{\thefigure}{S\arabic{figure}}

%\tableofcontents

%\newpage

\subsection{Heterostructure}

The vdW MoSe$_2$/WSe$_2$ HS (Fig.~S1a) was assembled using the dry-transfer peel technique~\cite{Withers2015}. The HS manufacturing details are described in Ref.~\cite{Fowler-Gerace2024}, where the same HS was used for cw studies of IXs. The thickness of the bottom hBN layer is $\sim 40$~nm, the thickness of the top hBN layer is $\sim 30$~nm. The MoSe$_2$ monolayer is on top of the WSe$_2$ monolayer. The long WSe$_2$ and MoSe$_2$ edges (Fig.~S1b) enable a rotational alignment between the WSe$_2$ and MoSe$_2$ monolayers. The twist angle between the monolayers $\delta \theta = 1.1^\circ$ corresponding to the moir{\'e} superlattice period $b = 17$~nm agrees with the angle between MoSe$_2$ and WSe$_2$ edges in the HS (Fig.~S1b). 

The accuracies of estimating  $\delta \theta$ using the long WSe$_2$ and MoSe$_2$ edges and using SHG are comparable. We do not use SHG for additional estimates of $\delta \theta$ since the intense excitation pulses in SHG measurements may cause a deterioration of the HS and may suppress the efficient IX transport. The moir{\'e} potentials can be affected by atomic reconstruction~\cite{Weston2020, Rosenberger2020, Zhao2023} and by disorder and may vary over the HS area. 

The IX $g$-factor is determined by the HS stacking~\cite{Seyler2019, Wozniak2020}. The IX $g$-factor $\sim - 16$ in the MoSe$_2$/WSe$_2$ HS corresponds to H stacking.

Figure~S1b presents a microscope image showing the layer pattern of the HS. The layer boundaries are indicated. The hBN layers cover the entire areas of MoSe$_2$ and WSe$_2$ layers. There was a narrow multilayer graphene electrode on the top of the HS around $x = 2$~$\mu$m for $y = 0$ in Fig.~S1b. This electrode was detached. 

So far, the efficient IX transport kinetics in vdW TMD HS was realized in one sample in this work. Other samples in other studies of vdW TMD heterostructures show diffusive IX transport characterized 
by significantly smaller diffusivities as outlined in the main text. The lack of the efficient IX transport kinetics likely originates from stronger disorder, which causes IX scattering and localization. When disorder is large it breaks exciton superfluidity. With lowering the disorder, superfluidity first emerges in small disconnected puddles, then the puddles grow in size, and for the sufficiently low disorder superfluid percolating through the entire sample emerges. This leads to efficient IX transport through the entire vdW TMD HS. This work demonstrates the existence of the efficient IX transport kinetics in TMD heterostructures. It is essential to study this phenomenon in other samples with different HS parameters. However, the sample statistics of other IX transport studies in vdW TMD heterostructures outlined in the main text shows that it is challenging to manufacture samples with different HS parameters, all with sufficiently small disorder. This is the subject for future works.

\begin{figure}
\begin{center}
\includegraphics[width=10cm]{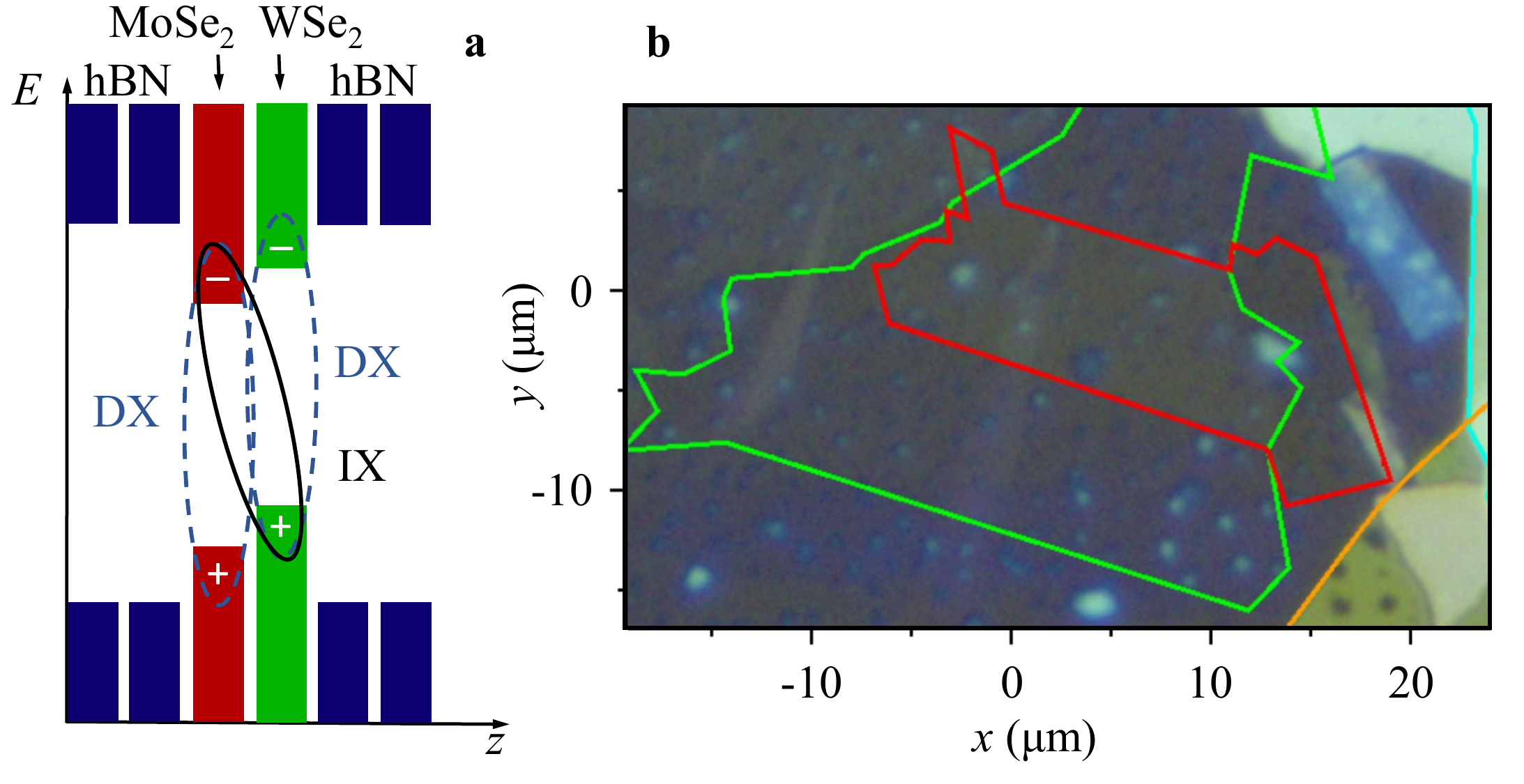}
\caption{(a) Schematic energy-band diagram of the MoSe$_2$/WSe$_2$ vdW HS. The ovals indicate DXs and IX composed of an electron (--) and a hole (+).
(b) A microscope image showing the HS layers. The green, red, cyan, and orange lines indicate the boundaries of WSe$_2$ and MoSe$_2$ monolayers and bottom and top hBN layers, respectively. 
}
\end{center}
\label{fig:spectra}
\end{figure}

\subsection{Optical measurements}

Excitons were generated by a semiconductor laser with the excitation energy $E_{\rm ex} = 1.689$~eV resonant to DX in WSe$_2$ HS layer. The laser pulses had step-like rectangular shape with 84~ns pulse width, 0.1~ns pulse rise time, 0.3~ns pulse fall time (the pulses rise and fall $e$ times within 0.1 and 0.3~ns, respectively), and 300~ns pulse period. The 
$\sim~220$~ns off time exceeded the IX lifetime (Fig.~S3) and was sufficient for a substantial decay of the IX signal.

The IX transport was measured by time- and spatially-resolved optical imaging using an avalanche photodiode (APD). The APD was moved in the plane of the optical image to measure the IX PL kinetics $I(x, t)$. For the image magnification x$92$ and the APD window diameter 100~$\mu$m, the spatial integration of the imaging setup $\sim 1.1$~$\mu$m was substantially smaller than the IX transport distances (Fig.~1 in the main text). The APD time resolution was 0.1~ns. Both the spatial and time resolution of the time-resolved imaging setup was sufficient for the time-resolved imaging of the IX transport. The IX emission in the time-resolved imaging experiments was spectrally selected by the long-pass filter with cut-off wavelength 850~nm (1.459~eV). 

PL spectra were measured using a spectrometer with a resolution of 0.2~meV and a liquid-nitrogen-cooled charge-coupled device. The experiments were performed in a variable-temperature 4He cryostat. The sample was mounted on an Attocube $xyz$ piezo translation stage allowing adjusting the sample position relative to a focusing lens inside the cryostat. All phenomena presented in this work are reproducible after multiple cooling down to 2~K and warming up to room temperature.

\subsection{Spatial profiles $I(x, t)$} 

A representative example of IX cloud expansion with time $I(x, t)$ in the regime of efficient IX transport is presented in Fig.~1a in the main text. Figure~S2 shows a representative example of $I(x, t)$ in the regime of regular diffusive IX transport. In contrast to the efficient IX transport with the fast enhancement of the spatial extension of IX cloud $d_{1/e}(t)$ with time (Fig.~1a in the main text), for diffusive IX transport the enhancement of $d_{1/e}(t)$ with time is significantly slower (Fig.~S2). The difference between the efficient IX transport regime and the diffusive transport regime is characterized, in particular, by the anomalously high IX diffusivity for the former and is outlined in the main text.

\begin{figure}
\begin{center}
\includegraphics[width=5.5cm]{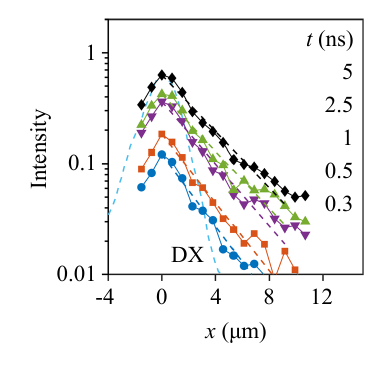}
\caption{IX PL spatial profiles $I(x)$ vs. time $t$ after the start of laser excitation pulse. IXs are created by the laser excitation. The excitation spot is centered at $x = 0$ and is shown by cyan dotted line that presents the DX PL profile in MoSe$_2$ monolayer and is close to the laser profile for short-range DX transport. The dashed lines show least-squares fits of the IX cloud profiles $I(x, t)$ to exponential decays in the region from the excitation spot to the HS edge, $x = 0 - 9$~$\mu$m. $P_{\rm ex} = 0.5$~mW, $T = 50$~K. The IX cloud expands and the $1/e$ decay distance of the IX cloud $d_{1/e}$ increases with time due to IX transport. In contrast to the data in Fig.~1a in the main text presenting $I(x, t)$ for the efficient IX transport kinetics, the enhancement of $d_{1/e}$ with time in Fig.~S2 is significantly slower that is characterized, in particular, by significantly lower IX diffusivity at these parameters. The diffusivity dependence on parameters is presented in Fig.~3 in the main text.
}
\end{center}
\label{fig:spectra}
\end{figure}

\subsection{Lifetime}

The IX recombination lifetime $\tau$ was obtained from the measurements of the decay kinetics of the spatially integrated IX signal after the excitation pulse end, $I_{\rm int}(t)$. The spatial integration minimizes the effect of IX transport on the measured kinetics and realizes the regime where the decay kinetics is governed by the IX recombination. Both increasing excitation power and temperature led to a reduction of $\tau$ (Fig.~S3). These variations of $\tau$ were included in the simulations using Eqs.~1, 2, and 3 outlined in the main text, where the measured values of $\tau(P_{\rm ex}, T)$ corresponding to the considered $T$ and $P_{\rm ex}$ were used. The recombination comprises all recombination processes including Auger recombination. 

The recombination rates, given by the slope of $\ln \left ( I_{\rm int} (t) \right )$, reduce with time (Fig.~S3a-c). The initial recombination rate after the excitation pulse end, $\tau^{-1}$, corresponds to the IX density during the pulse. Since IX transport during the pulse is considered in the experiments (Fig.~2 in the main text), the measured initial lifetime $\tau$ (Fig.~S3d) is used in the simulations.
For the efficient IX transport regime, the effect of lifetime variation on kinetics is rather weak and we checked that this variation leads to minor $\sim 10 \%$ changes of the parameters obtained from the kinetics fits.

\begin{figure}
\begin{center}
\includegraphics[width=8.5cm]{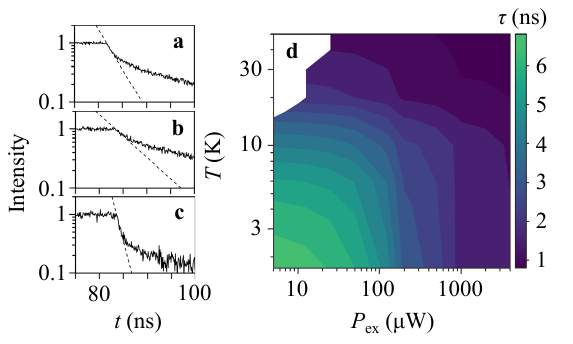}
\caption{(a-c) The decay kinetics of the spatially integrated IX signal after the excitation pulse end, $I_{\rm int}(t)$, for different excitation powers and temperatures: $P_{\rm ex} = 200$~$\mu$W, $T = 1.7$~K in (a), $P_{\rm ex} = 50$~$\mu$W, $T = 1.7$~K in (b), $P_{\rm ex} = 200$~$\mu$W, $T = 20$~K in (c). The dashed lines showing the initial slope of $\ln \left ( I_{\rm int} (t) \right )$ after the excitation pulse end at $t \sim 84$~ns are used to estimate $\tau$. (d)
The IX recombination lifetime $\tau$ vs. excitation power $P_{\rm ex}$ and temperature $T$. 
$\tau$ is the initial decay time after the excitation pulse end.
}
\end{center}
\label{fig:spectra}
\end{figure}

\subsection{Simulated spatial profiles $I(x, t)$} 

Examples of simulated spatial profiles $I(x, t)$ for IX diffusion and ballistic transport using, respectively, Eq.~1 and Eq.~2 in the main text are shown in Fig.~S4. The simulations used the spatial profile of the laser excitation spot $\Lambda(x)$, given by the measured DX PL profile in MoSe$_2$ monolayer that is close to the laser profile for short-range DX transport, and the IX lifetime $\tau$ presented in Fig.~S3.

\begin{figure}
\begin{center}
\includegraphics[width=10cm]{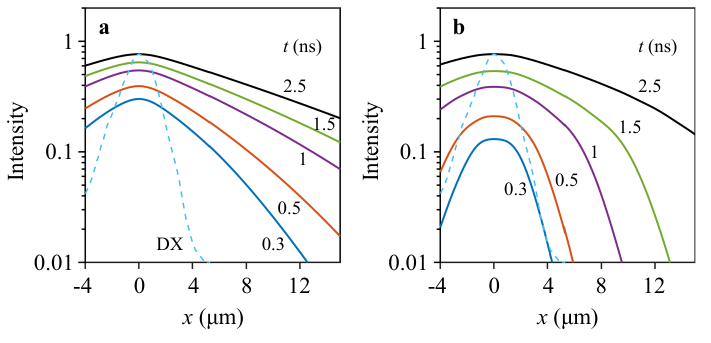}
\caption{Simulated spatial profiles $I(x, t)$ for IX diffusion (a) and ballistic transport (b) using, respectively, Eq.~1 and Eq.~2 in the main text. For these simulations, the IX diffusivity $D = 720$~cm$^2$/s in (a), the IX velocity $v = 7.4 \cdot 10^5$~cm/s in (b), and the IX lifetime $\tau = 2.7$~ns in (a,b). The dotted lines show the spatial profile of the laser excitation spot $\Lambda(x)$. The excitation pulse starts at $t = 0$. 
}
\end{center}
\label{fig:spectra}
\end{figure}

The mean-squared displacement (MSD) and $d_{1/e}$, which quantify the spatial extension of IX spatial profiles $I(x, t)$, grow with time. MSD is close to $d_{1/e}^2$ as shown by Fig.~S5. The simulations for diffusion (Fig.~2a) show nearly linear growth of MSD and $d_{1/e}^2$ with time, or, equivalently, nearly square-root-power growth of $d_{1/e}$ with time. The simulations for ballistic transport (Fig.~2b) 
and for drift-diffusion transport (Fig.~2c) 
show nearly quadratic growth of MSD and $d_{1/e}^2$ with time, or, equivalently, nearly linear growth of $d_{1/e}$ with time. 

There are deviations from linear growth of MSD and $d_{1/e}^2$ with time for diffusion and there are deviations from quadratic growth of MSD and $d_{1/e}^2$ with time for ballistic transport 
and drift-diffusion transport. In particular, exciton recombination makes the growth slower. 

While $d_{1/e}^2$ and MSD are close (Fig.~S5), using $d_{1/e}^2$ (or $d_{1/e}$) is more accurate for describing experimental data. In particular, in experiments, any sample has a finite size and $d_{1/e}$ accurately quantifies the expansion of spatial profiles $I(x, t)$ even when $d_{1/e}$ exceeds the sample size, however, MSD is affected by the finite sample size and cannot be used in this case.

\begin{figure}
\begin{center}
\includegraphics[width=4.5cm]{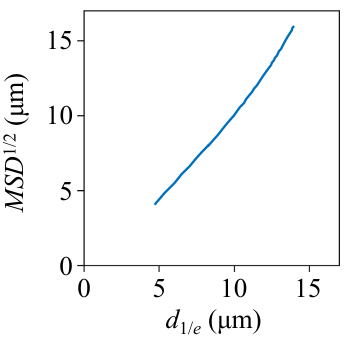}
\caption{Comparison of MSD and $d_{1/e}$ for $I(x, t)$ in Fig.~S4a. $d_{1/e}^2$ and MSD are close.
}
\end{center}
\label{fig:spectra}
\end{figure}

\subsection{On analysis of $d_{1/e}(t)$}

In the main text, the IX diffusivity $D$ is estimated by least-squares fit of the kinetics of exciton cloud expansion $d_{1/e}(t)$ simulated with Eq.~1 to the measured $d_{1/e}(t)$ in the time range $t = 0 - 5$~ns (Fig.~2a), and the IX velocity $v$ is estimated by least-squares fit of the kinetics of exciton cloud expansion $d_{1/e}(t)$ simulated with Eq.~2 to the measured $d_{1/e}(t)$ in the time range $t = 0 - 5$~ns (Fig.~2b). The results are similar for the fits made in different time ranges. Figure~S6 shows the fits similar to the fits in Fig.~2 in the main text, however, for the fit range $t = 0 - 3$~ns. These fits give the values of $D$ and $v$ shown in Fig.~S7. The fits over $t = 0 - 3$~ns give similar $D$ and $v$ values and their similar temperature and excitation density dependence $D(P_{\rm ex}, T)$ and $v(P_{\rm ex}, T)$ as the fits over $t = 0 - 5$~ns in the main text (compare Fig.~S7 and Fig.~3 in the main text). Furthermore, for the efficient IX transport, the fits over $t = 0 - 3$~ns show a high deviation of the measured $d_{1/e}(t)$ from diffusion (Fig.~S6a) and significantly smaller deviation of the measured $d_{1/e}(t)$ from ballistic transport (Fig.~S6b), similarly to the fits over $t = 0 - 5$~ns as outlined in the main text. The specifics of the fits over $t = 0 - 3$~ns is that $d_{1/e}(t)$ in this time range is within the HS area (Fig.~S6).

The estimates are obtained using Eqs.~1 and 2 with constant values for $D$ and $v$. During the excitation pulse, the IX density varies with time. Simulations of IX kinetics with density-dependent parameters can improve the accuracy of estimates and is the subject of future works. 

\begin{figure}
\begin{center}
\includegraphics[width=8.5cm]{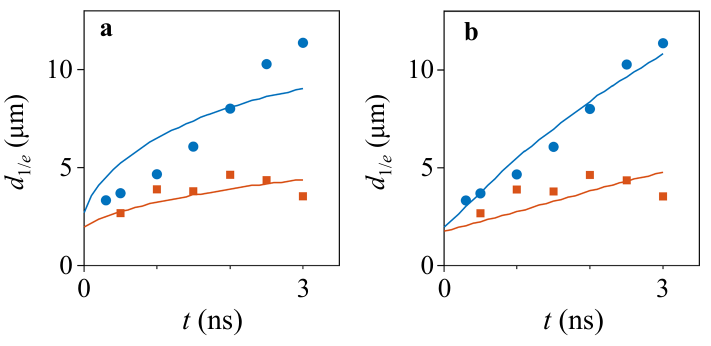}
\caption{Simulations of IX cloud expansion kinetics. 
The measured IX cloud expansion $d_{1/e}(t)$ is shown by blue points for the excitation power $P_{\rm ex} = 500$~$\mu$W and by red squares for $P_{\rm ex} = 25$~$\mu$W. The simulated $d_{1/e}(t)$ for IX diffusion (a) and ballistic transport (b) are shown by lines. In the simulations, the IX diffusivity $D$ for diffusion (a) and IX velocity $v$ for ballistic transport (b) are estimated by least-squares fit of the kinetics of exciton cloud expansion $d_{1/e}(t)$ simulated with Eq.~1 and Eq.~2 in the main text, respectively, to the measured $d_{1/e}(t)$. The simulations in Fig.~S6 are similar to the simulations in Fig.~2 in the main text, however, the fit range is $t = 0 - 3$~ns in Fig.~S6 and $t = 0 - 5$~ns in Fig.~2 in the main text. The estimates for $D$ and $v$ obtained with the simulations with the fit range $t = 0 - 3$~ns (Fig.~S6) are presented in Fig.~S7. The laser excitation pulse starts at $t = 0$. $T = 1.7$~K.
}
\end{center}
\label{fig:spectra}
\end{figure}

\begin{figure}
\begin{center}
\includegraphics[width=8.5cm]{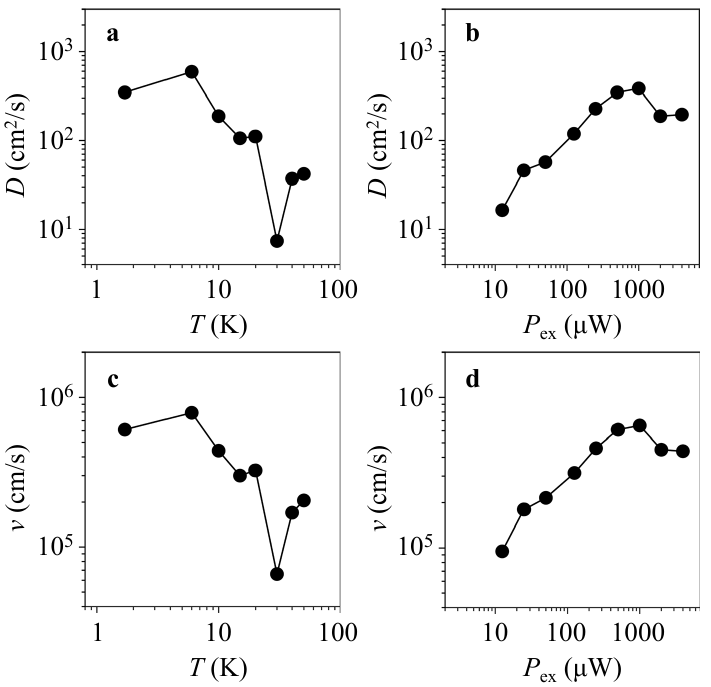}
\caption{Temperature and excitation power dependence of IX transport kinetics: 
diffusion approach (a,b) and ballistic transport approach (c,d). The IX diffusivity for diffusion and IX velocity for ballistic transport are estimated by least-squares fit of the kinetics of exciton cloud expansion $d_{1/e}(t)$ simulated with Eq.~1 and Eq~2 in the main text, respectively, to the measured $d_{1/e}(t)$. The estimates in Fig.~S7 are similar to the estimates in Fig.~3 in the main text, however, the fit range is $t = 0 - 3$~ns in Fig.~S7 and $t = 0 - 5$~ns in Fig.~3 in the main text. 
(a,b) The estimated IX diffusivity $D$ vs. temperature (a) and excitation power (b). 
(c,d) The estimated IX velocity $v$ vs. temperature (c) and excitation power (d). 
$P_{\rm ex} = 0.5$~mW (a,c), $T = 1.7$~K (b,d).
}
\end{center}
\label{fig:spectra}
\end{figure}

\begin{figure}
\begin{center}
\includegraphics[width=4cm]{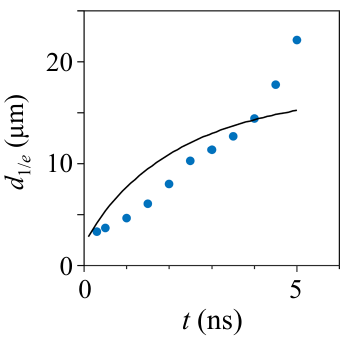}
\caption{Simulations of IX cloud expansion kinetics due to diffusion and interaction-induced drift. The measured IX cloud expansion $d_{1/e}(t)$ is shown by blue points for $P_{\rm ex} = 500$~$\mu$W and $T = 1.7$~K. The simulated $d_{1/e}(t)$ for IX transport due to diffusion and interaction-induced drift is shown by line. The simulations use Eq.~1 with the additional term describing the interaction-induced drift as outlined in the main text and SI. The laser excitation pulse starts at $t = 0$.
}
\end{center}
\label{fig:spectra}
\end{figure}

\subsection{On long-distance IX transport and IX mediated long-distance spin transport}

The density and temperature range of the efficient IX transport kinetics 
characterized by 
the anomalously high diffusivity (Fig.~3 in the main text) and nearly quadratic growth of $d_{1/e}^2$ with time, or, equivalently, nearly linear growth of $d_{1/e}$ with time (Fig.~2 in the main text), is consistent with the density and temperature range for the long-distance IX transport and the IX mediated long-distance spin transport in cw studies of IXs in vdW TMD HS~\cite{Fowler-Gerace2024, Zhou2024}. This indicates that the latter phenomena originate from the efficient IX transport kinetics.

\subsection{IX interaction} 

The IX out-of-plane dipoles cause repulsive interaction between IXs and, as shown in Ref.~\cite{Yoshioka1990}, the 
repulsive interaction between IXs can be approximated by the mean-field formula for the IX energy shift with density $\delta E = n u_0$, $u_0 = 4 \pi e^2 d_z/\varepsilon$, where $d_z$ is the separation between the electron and hole layers and $\varepsilon$ is the dielectric constant ($d_z \sim 0.6$~nm and $\varepsilon \sim 7.4$ for the HS~\cite{Laturia2018}). 
Including correlations beyond this mean-field model lowers $u_0$ in comparison to the mean-field value, leading to a higher estimated $n$ for a given $\delta E$ as outlined in the main text. 
The IX repulsive interaction introduces interaction-induced drift from the origin 
adding the term $\nabla \left[ \mu n \nabla (n u_0) \right]$ to Eq.~1~\cite{Ivanov2002}. For the IX mobility $\mu$ estimated by the Einstein relation $\mu = D_0/k_BT$, IX diffusion with interaction-induced drift can be approximately described as diffusion with the enhanced diffusivity $D = D_0 (1 + nu_0 / k_{\rm B} T)$, where $D_0$ is the diffusivity for non-interacting particles, as outlined in Ref.~\cite{Dorow2017}. Similar to diffusion (Fig.~2a in the main text), $d_{1/e}^2(t)$ grows nearly linearly with time ($d_{1/e}(t)$ grows nearly $\propto t^{1/2}$) for IX diffusion with interaction-induced drift (Fig.~S8). 

For the efficient IX transport kinetics at $P_{\rm ex} \sim 500$~$\mu$W (Fig.~2 in the main text), the energy shift $\delta E \sim 3$~meV (Fig.~S11c). For this energy shift, 
the mean-field estimate for the IX density~\cite{Yoshioka1990} outlined in the main text gives $n \sim 2 \cdot 10^{11}$~cm$^{-2}$. 

Alternatively, IX density can be roughly estimated from the light absorption $n \sim \alpha P_{\rm ex} \tau / E_{\rm ex} A_{\rm ex}$, where $A_{\rm ex}$ is the excitation area and $\alpha$ is the fraction of photons provided by the excitation laser that is transformed to IXs due to the light absorption and carrier relaxation. For $P_{\rm ex} \sim 500$~$\mu$W for the laser excitation pulses, $E_{\rm ex} = 1.689$~eV, $A_{\rm ex} \sim 150$~$\mu$m$^2$ for the IX expansion over the entire HS at this $P_{\rm ex}$ (Fig.~1a), $\alpha \sim 0.1$, and $\tau \sim 2.7$~ns for this $P_{\rm ex}$ (Fig.~S3), the estimate from the light absorption gives $n \sim 3 \cdot 10^{11}$~cm$^{-2}$. This value is close to the mean-field estimate from the IX energy shift outlined above. However, the exciton density estimates based on the light absorption may be inaccurate. Therefore, we use the mean-field estimate for the IX density based on the IX energy shift~\cite{Yoshioka1990} as outlined in the main text.

\begin{figure}
\begin{center}
\includegraphics[width=12.5cm]{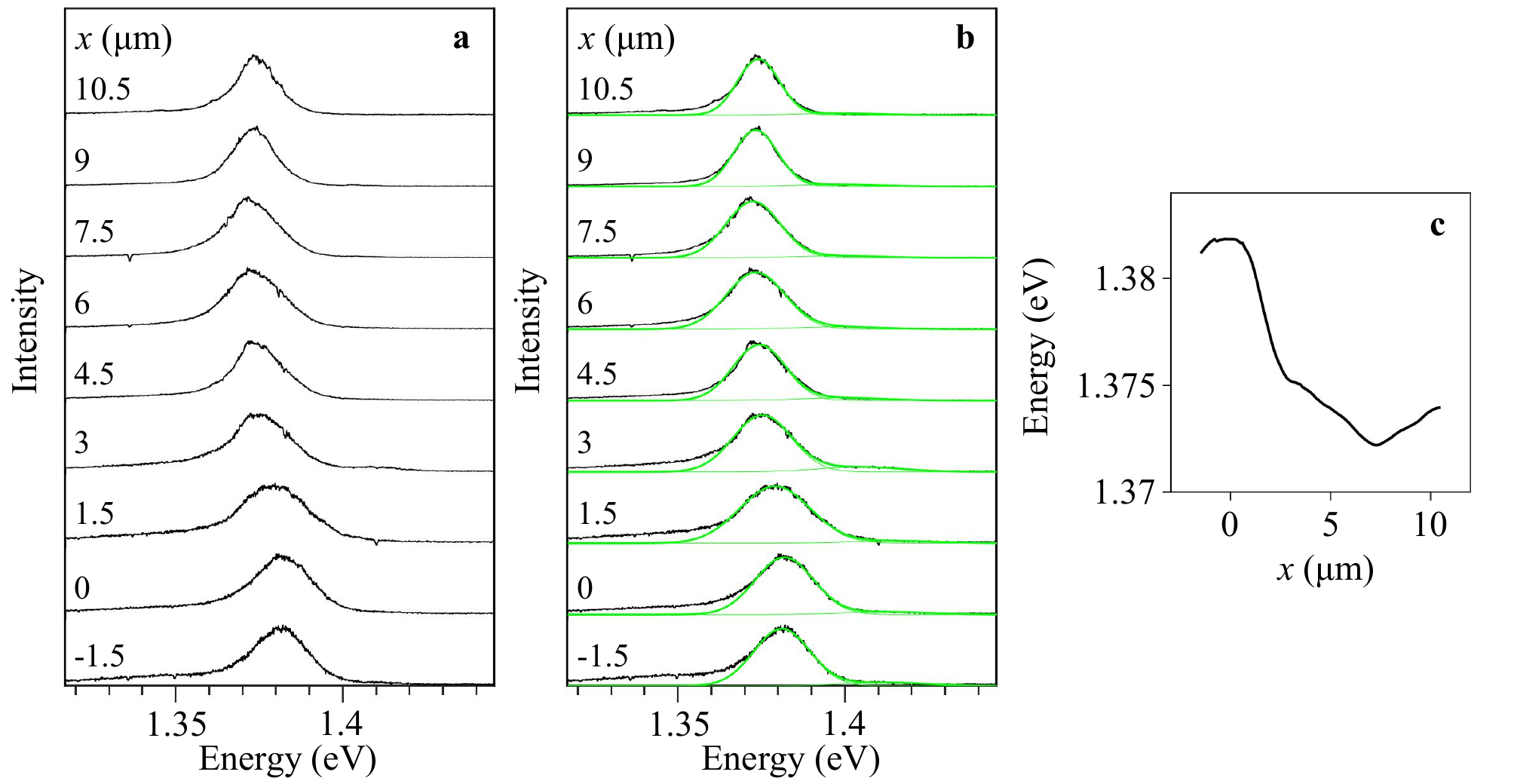}
\caption{Potential energy landscape. (a) IX spectra vs. position $x$ along the IX path. (b) Same spectra with the gaussian fits shown by the green lines. (c) Position dependence of the IX energy $U(x)$. The laser excitation is defocused to $\sim 25$~$\mu$m spot covering the entire HS, the power of this defocused excitation is $\sim 5$~mW. $T = 4.2$~K.
}
\end{center}
\label{fig:spectra}
\end{figure}

\subsection{IX drift due to potential energy landscape}

As outlined in the main text, Eq.~3 includes the term $\nabla \left[ \mu n \nabla U \right]$ describing IX drift due to the in-plane potential energy landscape in the HS, $U$. This landscape originates from the variation of the local IX environment in the HS. The specific origin of this landscape is not essential for the IX transport modeling by Eq.~3. The measured $U(x)$ in the HS is presented in Fig.~S9c. This figure shows that $U(x)$ drops $\sim 10$~meV on $\sim 10$~$\mu$m of the IX path from the excitation spot to the HS edge and, for the simulations using Eq.~3, $U(x)$ is approximated by the energy gradient 1~meV/$\mu$m. 

The simulations in Fig.~S8 include IX drift due to the IX interaction and do not include IX drift due to the in-plane potential energy landscape. These simulations show nearly square-root growth of $d_{1/e}$ with time and do not agree with the data showing nearly linear growth of $d_{1/e}$ with time. In contrast, the simulations including both drift due to the IX interaction and drift due to the in-plane potential energy landscape in the HS show nearly linear growth of $d_{1/e}$ with time (Fig. 2c) and agree with the data.

\subsection{Deviation of IX transport from diffusion, ballistic, and drift-diffusion transport}

We compared the deviation of IX transport kinetics from the simulated kinetics of diffusion (Fig.~2a), ballistic (Fig.~2b), and drift-diffusion (Fig.~2c) transport. The deviation is quantified by $\chi^2 = \Sigma (m_t - s_t)^2/(\nu s_t)$, where $m_t$ and $s_t$ are the measured and simulated IX cloud extension $d_{1/e}$ at time $t$ and $\nu$ is the number of the measured $t$ values in $\chi^2$ (Fig.~S10). In the regime of the efficient IX transport (yellow region in Figs.~3a3, 3b3, and 3c3), $\chi^2$ are large for diffusion (Figs.~S10a,b,c) and substantially smaller for the ballistic and drift-diffusion transport (Figs.~S10a,b,d,e). This confirms that the efficient IX transport kinetics significantly deviates from the simulated kinetics of diffusion and is close to the simulated kinetics of both drift-diffusion and ballistic transport. 

This $\chi^2$ test justifies the same conclusion, which can be made by the eye comparison of the measured and simulated $d_{1/e}(t)$ in Fig.~2: The data show a strong deviation of the measured efficient IX transport kinetics from the simulated kinetics of diffusion transport (the blue dots strongly deviate from the blue line in Fig.~2a) and significantly smaller deviation of the measured efficient IX transport kinetics from the simulated kinetics of both drift-diffusion and ballistic transport (the deviation of the blue dots from the blue line in Figs.~2b,c is significantly smaller).

\begin{figure*}
\begin{center}
\includegraphics[width=17.9cm]{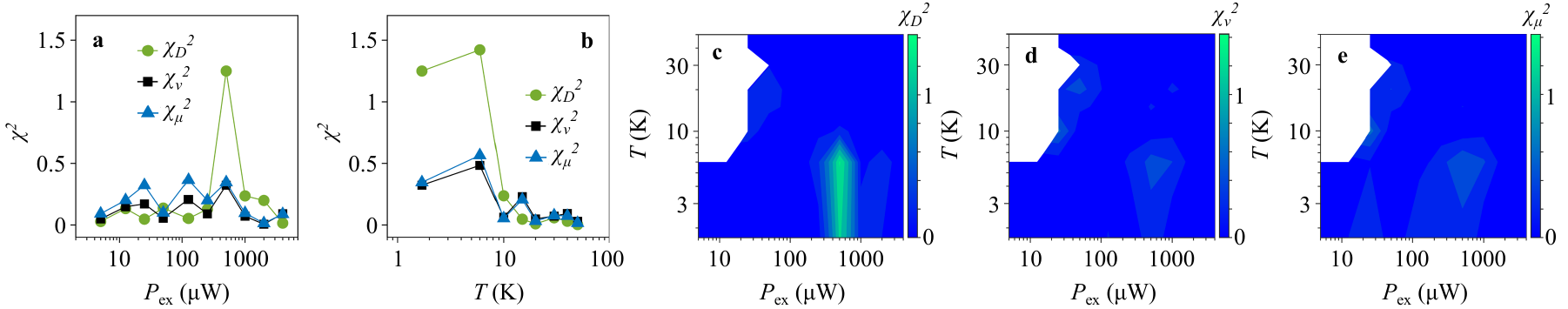}
\caption{Comparison of IX transport with diffusion, ballistic, and drift-diffusion transport.
(a-e) The deviation of the measured IX transport kinetics from the simulated kinetics $\chi^2 = \Sigma (m_t - s_t)^2/(\nu s_t)$ vs. excitation power (a), temperature (b), and excitation power and temperature (c-e). $m_t$ and $s_t$ are the measured and simulated IX cloud extension $d_{1/e}$ at time $t$ and $\nu$ is the number of the measured $t$ values in $\chi^2$. 
(a,b) Green points, black squares, and blue triangles show $\chi^2$ deviation of the measured IX transport from the simulated diffusion, ballistic, and drift-diffusion transport, respectively. 
(c-e) $\chi^2$ deviation of the measured IX transport from the simulated diffusion (c), ballistic (d), and drift-diffusion (e) transport. In the regime of the efficient IX transport (yellow region in Figs.~3a3, 3b3, and 3c3), $\chi^2$ are large for diffusion and significantly smaller for ballistic and drift-diffusion transport. This shows that the efficient IX transport kinetics significantly deviates from the kinetics of diffusion and is close to the kinetics of both ballistic transport and drift-diffusion transport.
$T = 1.7$~K (a), $P_{\rm ex} = 0.5$~mW (b).
}
\end{center}
\label{fig:spectra}
\end{figure*}

\subsection{Comparison with BH theory}

As outlined in the main text, the BH model predicts the superfluid phase and the efficient IX transport kinetics, characterized by the anomalously high IX mobility despite the in-plane disorder, is qualitatively consistent with the predicted superfluidity. Below, we discuss a rough comparison of the parameters, at which the efficient IX transport kinetics is observed, with the BH model. 

The BH model predicts the superfluid phase for the number of bosons per lattice site $N \sim 1/2$ and the insulating phase, such as the Mott insulator and the Bose glass, for $N \sim 0$ and $N \sim 1$~\cite{Fisher1989}. 
The IX transport enhancement followed by the suppression with density (Fig.~3 in the main text) is qualitatively consistent with this prediction. The rough estimates also indicate a qualitative agreement with the BH model: For the efficient IX transport kinetics at $P_{\rm ex} \sim 500$~$\mu$W (Fig.~2 in the main text), the energy shift $\delta E \sim 3$~meV (Fig.~S11c); 
For this energy shift, 
the mean-field estimate for the IX density, outlined in SI Section {\it IX interaction}, gives $n \sim 2 \cdot 10^{11}$~cm$^{-2}$; $N \sim 1/2$ for this density and the moir{\'e} superlattice period $b = 17$~nm, where $b \sim a/\delta \theta$ corresponds to the twist angle $\delta \theta = 1.1^\circ$ that agrees with the angle between MoSe$_2$ and WSe$_2$ layers in the HS ($a$ is the lattice constant).

The efficient IX transport is observed in a broad range of excitation powers $P_{\rm ex}$ (Fig.~3) and, in turn, $\delta E$ given by $P_{\rm ex}$ (Fig.~S11c) and densities $n$ estimated from $\delta E$ as outlined in SI Section {\it IX interaction}. This indicates that the efficient IX transport is observed in a broad range of $N$. The BH model also predicts superfluidity in a broad range of $N$~\cite{Fisher1989}. The density estimates in this work are performed within the mean-field model~\cite{Yoshioka1990} and these rough estimates give $N$ for the efficient IX transport qualitatively consistent with $N$ predicted for superfluidity in the Bose-Hubbard model. The broad density ranges of both the observed efficient IX transport and superfluidity in the BH model indicate that the density estimate does not have to be precise for this qualitative consistency. 

The efficient IX transport vanishes above $T \sim 10$~K (Fig.~3). 
As outlined in the main text this is inconsistent with diffusion and, in contrast, is qualitatively consistent with superfluidity. The rough estimates, presented below, also indicate a qualitative agreement with the BH model: 
The BH theory predicts the critical temperature for superfluidity $T_{\rm c} \sim 4 \pi N J$~\cite{Capogrosso-Sansone2008}; 
For $T_{\rm c} \sim 10$~K 
this formula gives 
the intersite hopping $J \sim 1.6$~K and, in turn, the moir{\'e} superlattice amplitude $\sim 8$~meV for $b \sim 17$~nm~\cite{Remeika2012}; This value is comparable to the moir{\'e} potential amplitude for the MoSe$_2$/WSe$_2$ HS with H stacking~\cite{Wu2018, Yu2018, Wu2017, Yu2017} and the $g$-factor in the HS corresponds to H stacking~\cite{Seyler2019, Wozniak2020}.

\begin{figure}
\begin{center}
\includegraphics[width=12.5cm]{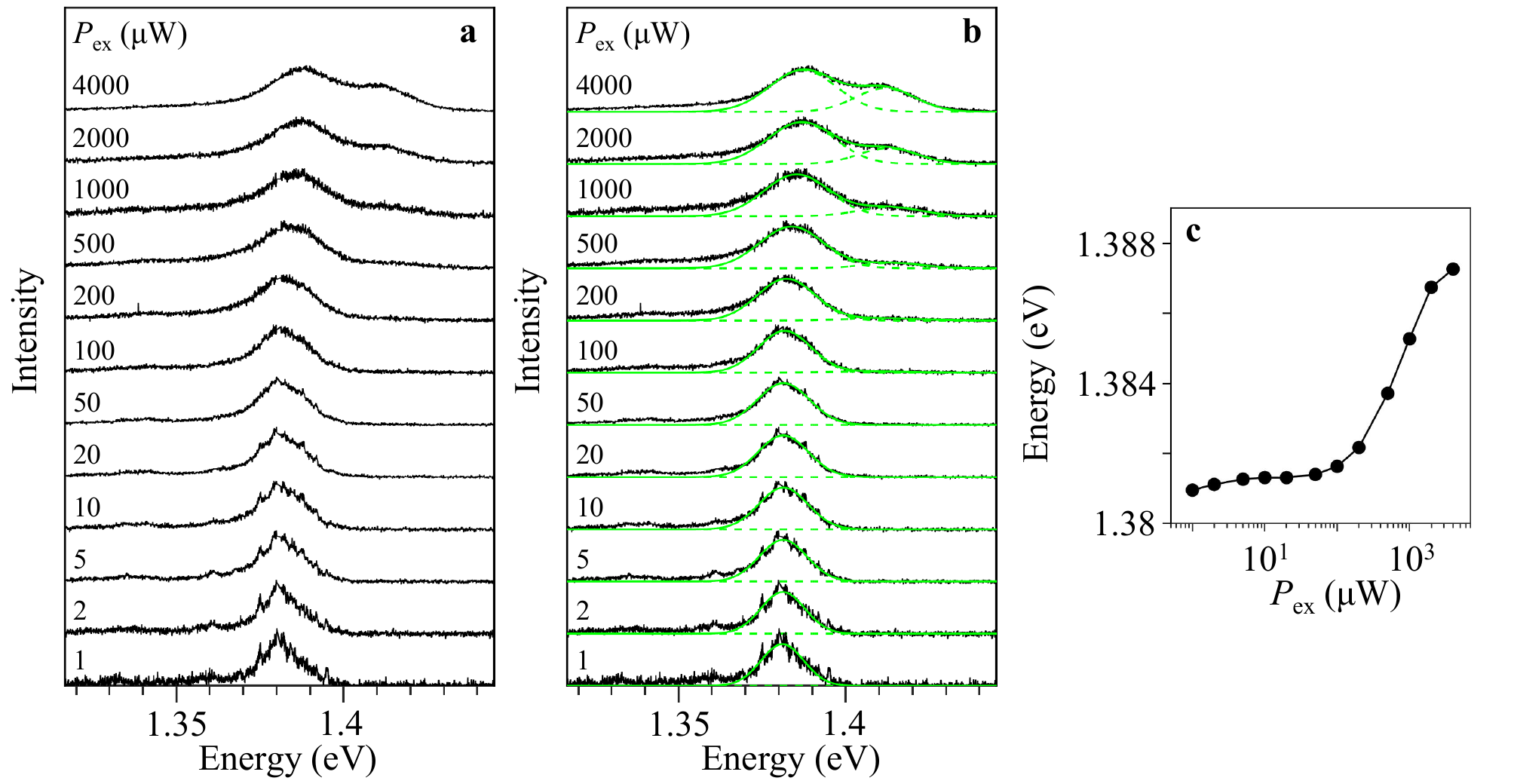}
\caption{IX PL spectra. (a) The excitation power $P_{\rm ex}$ dependence of IX spectra. $T = 6$~K. (b) Same spectra with the gaussian fits shown by the dashed green lines. The sum of the gaussians shown by the dashed green lines is shown by the solid green line. (c) $P_{\rm ex}$ dependence of the IX energy.
}
\end{center}
\label{fig:spectra}
\end{figure}

\subsection{IX PL spectra}

Figure~S11 shows IX spectra at different excitation powers $P_{\rm ex}$. The IX PL linewidth at low densities gives an estimate for the disorder amplitude in the HS. Both the smallest IX linewidth at the lowest densities $\sim 14$~meV and the IX linewidth at the excitation densities corresponding to the efficient IX transport $\sim 20$~meV (Fig.~S11a,b) are significantly larger than the IX interaction energy at the densities where the efficient IX transport is observed $\sim 3$~meV (Fig.~S11c). Therefore, the IX interaction energy is insufficient for effective screening of the in-plane potential for diffusive IX transport~\cite{Remeika2009}. The efficient IX transport kinetics occurs despite the strong in-plane potential that is inconsistent with classical diffusive transport as outlined in the main text. 

Narrow lines appearing in the IX spectrum at lower densities (Fig.~S11) were attributed to localized states~\cite{Zhou2024}. A higher-energy IX line appearing at high IX densities (Fig.~S11) was attributed to the appearance of moir{\'e} cells with double occupancy~\cite{Zhou2024}. The transport kinetics data are spectrally integrated over the entire IX PL and all IX states are included in the transport measurements.

The shift $\delta E$ of the lower-energy IX line with $P_{\rm ex}$ is shown in Fig.~S11c. This shift $\delta E$ is used for the mean-field estimate of IX density following Ref.~\cite{Yoshioka1990} as outlined in SI Section {\it IX interaction}. 

Figure~S11 shows that in the regime of the efficient IX transport kinetics ($P_{\rm ex} \sim 500$~$\mu$W) and up to the highest densities the IX linewidth is close to the linewidth at the lowest densities. This indicates the excitonic regime for the studied IX transport kinetics. The excitonic regime is consistent with the estimated IX density $n \sim 2 \cdot 10^{11}$~cm$^{-2}$ for the efficient IX transport, significantly smaller than the density of the Mott transition to electron-hole plasma $n_{\rm Mott} > 10^{12}$~cm$^{-2}$~\cite{Fogler2014, Wang2019}. In contrast, in the plasma regime, the linewidth substantially increases~\cite{Wang2019}.

\begin{figure}
\begin{center}
\includegraphics[width=10.5cm]{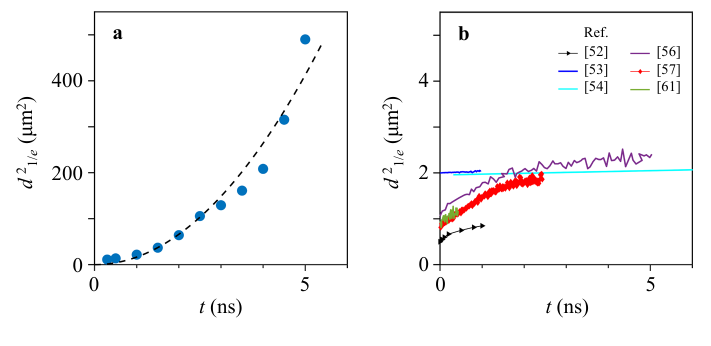}
\caption{Comparison of the efficient IX transport kinetics with earlier studies. 
(a) $d_{\rm 1/e}^2(t)$ for the efficient IX transport kinetics (same data for $d_{\rm 1/e}(t)$ are shown by blue dots in Fig.~2 in the main text). 
(b) $d_{\rm 1/e}^2(t)$ for IX transport kinetics in earlier studies of vdW TMD HS with Ref.~numbers indicating references in the main text. The $1/e$ decay distances of exciton signal $d_{\rm 1/e}$ are extracted from figures in these works for the data corresponding to the fastest IX transport kinetics in the excitonic regime in these works. A quadratic growth $d_{\rm 1/e}^2 \propto t^2$ is shown by dashed line for comparison with the data in (a).
}
\end{center}
\label{fig:spectra}
\end{figure}

\subsection{Comparison with earlier studies of IX transport kinetics in vdW heterostructures}

Figure~S12a shows $d_{\rm 1/e}^2(t)$ for the efficient IX transport kinetics (same data for $d_{\rm 1/e}(t)$ are shown by blue dots in Fig.~2 in the main text). Figure~S12b shows $d_{\rm 1/e}^2(t)$ for IX transport kinetics in earlier studies of vdW TMD HS with Ref.~numbers indicating references in the main text. The $1/e$ decay distances of exciton signal $d_{\rm 1/e}$ are extracted from figures in these works for the data corresponding to the fastest IX transport kinetics in the excitonic regime in these works. 
Figure S12 shows the following: (i) For the efficient IX transport kinetics (Fig.~S12a), $d_{\rm 1/e}$ grows significantly faster than for the IX transport kinetics in earlier studies (Fig.~S12b); (ii) In earlier studies (Fig.~S12b), $d_{\rm 1/e}^2$ grows linearly or sub-linearly with time that is characteristic of diffusion. In contrast, for the efficient IX transport kinetics (Fig.~S12a), $d_{\rm 1/e}^2$ grows nearly quadratically with time. 

A quantitative comparison of IX transport characteristics can be done by comparing the IX mean free time $\tau_{\rm m}$ estimated from the measured IX transport kinetics. For the efficient IX transport kinetics, the IX mobility $\mu \sim 7 \cdot 10^4$~cm$^2$/eVs at $T \sim 6$~K (Fig.~3c) gives the estimated $\tau_{\rm m} \sim \mu m \sim 40$~ps. In comparison, for the IX transport kinetics in Ref.~[56] in the main text (this work shows fastest IX transport kinetics among the earlier studies, see Fig.~S12b), the reported IX diffusivity $D \sim 0.15$~cm$^2$/s at $T = 4.6$~K gives the estimated $\tau_{\rm m} \sim D m / k_{\rm B} T \sim 0.2$~ps. The estimated IX mean free time for the efficient IX transport kinetics is significantly higher than in earlier studies. The differences between the efficient IX transport kinetics and transport kinetics in earlier studies are discussed in the main text.